\documentclass[twocolumn]{aastex62}
\usepackage[utf8]{inputenc}
\usepackage{natbib}
\usepackage{graphicx}
\usepackage{subfigure}

\usepackage{xcolor}

\shorttitle{External Enrichment of Minihalos}
\shortauthors{Hicks et al.}

\newcommand{\R}{M_{Z_3,p}/M_{Z_3,a}}
\newcommand{\logR}{\log_{10}\left(\R\right)}
\newcommand{\zpii}{Z_{2}}
\newcommand{\zpiii}{Z_{3}}

\begin{document}

\title{External Enrichment of Minihalos by the First Supernovae}

\author{William Hicks}
\email{whicks@ucsd.edu}
\affil{Center for Astrophysics and Space Sciences\\
University of California, San Diego, La Jolla, CA, 92093}

\author{Azton Wells}
\affil{Center for Astrophysics and Space Sciences\\
University of California, San Diego, La Jolla, CA, 92093}

\author{Michael L. Norman}
\affil{Center for Astrophysics and Space Sciences\\
University of California, San Diego, La Jolla, CA, 92093}
\affil{San Diego Supercomputer Center\\
University of California, San Diego, La Jolla, CA, 92093}
\author{John H. Wise}
\affil{Center for Relativistic Astrophysics, School of Physics\\
Georgia Institute of Technology, Atlanta, GA, 30332}
\author{Britton D. Smith}
\affil{Institute for Astronomy, University of Edinburgh, Royal Observatory Edinburgh, EH9 3HJ, UK}
\author{Brian W. O'shea}
\affil{Department of Physics and Astronomy, Michigan State University, East Lansing, MI 48824, USA}
\affil{ Department of Computational Mathematics, Science, and Engineering, Michigan State University, East Lansing, MI 48824, USA }
\affil{National Superconducting Cyclotron Laboratory, Michigan State University, East Lansing, MI 48824, USA}
\affil{ JINA-CEE: Joint Institute for Nuclear Astrophysics-Center for the Evolution of the Elements, USA }
\begin{abstract}
Recent high-resolution simulations of early structure formation have shown that externally enriched halos may form some of the first metal enriched stars. This study utilizes a 1 comoving Mpc$^3$ high-resolution simulation to study the enrichment process of metal-enriched halos down to $z=9.3$. Our simulation uniquely tracks the metals ejected from Population III stars, and we use this information to identify the origin of metals within metal-enriched halos. These halos show a wide range of metallicities, but we find that the source of metals for $\gtrsim$ 50\% of metal-enriched halos is supernova explosions of Population III stars occuring outside their virial radii. The results presented here indicate that external enrichment by metal-free stars dominates the enrichment process of halos with virial mass \textcolor{black}{below $10^{6}\,M_\odot$} down to $z=9.3$. Despite the prevalence of external enrichment in low mass halos, Pop II stars forming due to external enrichment are rare because of the small contribution of low-mass halos to the global star formation rate \textcolor{black}{combined with low metallicities towards the center of these halos resulting from metal ejecta from external sources mixing from the outside-in}. The enriched stars that do form through this process have absolute metallicities below $10^{-3}\,Z_\odot$. We also find that the fraction of externally enriched halos increases with time, $\sim 90\%$ of halos that are externally enriched have $M_\mathrm{vir} < 10^6\,M_\odot$, \textcolor{black}{and that pair-instability supernovae contribute the most to the enrichment of the IGM as a whole and are thus are the predominant supernova type contributing to the external enrichment of halos.}

\end{abstract}
\section{Introduction}
In a quest to discover the characteristics of the first-generation of stars, we inevitably run into the inconvenient fact that there are no observable first-generation (Population III, Pop III) stars, so direct constraints on their characteristics are lacking. This difficulty has given rise to galactic archaeology (i.e.,~\citealt{FELTZING201380, 2014Natur.506..463K,hernandez2020,frebel2015}) to determine the enrichment history of an object. Other observations attempt to constrain the Pop III initial mass function (IMF), number of enriching events and typical supernova energy~\citep{welsh2019}, or to detect the unique metal signature of massive Pop III supernovae~\citep{banados2019} by observing metal-poor damped Lyman-$\alpha$ (DLA) systems. In order to more fully understand the results from surveys measuring metallicity of presently observed stars or DLAs, it is imperative to understand how Pop III stars interacted with their environments. Prior works have investigated the effect of Pop III supernovae at the extreme small scale (\citealt{Whalen2008b, cen2008, doi:10.1093/mnras/stv1509, Chen2017}). These studies illustrate that pristine minihalos can be enriched by external sources, such as Type II supernovae (SNe) from Pop III stars forming in nearby minihalos. Notably, these externally enriched minihalos could be the first sites of second-generation star (Pop II) star formation~\citep{doi:10.1093/mnras/stv1509}. These prior simulations had extremely fine small-scale resolution, but were limited by the small simulation box size ($\sim 500$~comoving kpc/h) and the low number of halos present in the volume. Further, \cite{doi:10.1093/mnras/stv1509} only considered the effects of 40~M$_\odot$ Pop III stars, whereas a more comprehensive IMF would include a variety of stellar endpoints: Type II SNe, hypernovae (HNe), and pair-instability supernovae (PISN).  This study aims to extend \cite{doi:10.1093/mnras/stv1509} by making use of a simulation similar to that in Paper II of the \textit{Birth of a Galaxy} series \citep{wise2012b} that has a larger box size (1 comoving Mpc) and includes PISN, HNe, and black hole collapse as Pop III endpoints.

The remainder of this paper is organized as follows: Section \ref{sec:simulation} outlines the simulation design and reviews important stellar formation and feedback parameters.  Sections \ref{sec:enrichment} and \ref{sec:star_formation} present our analysis on the prospects of external enrichment and star formation that occurs afterwards. Finally, Section \ref{sec:discussion} compares our results to other studies and Section \ref{sec:conclusion} summarizes the main conclusions drawn from our analysis.

\section{Simulation Setup}
\label{sec:simulation}

The simulation used for this analysis is the same simulation analyzed in \cite{Skinner2020}. Making use of the AMR simulation code Enzo \citep{Enzo2014,Enzo2019}, the 1 Mpc box has a base resolution of $256^3$ cells and particles with up to 12 levels of local refinement, which results in a maximum comoving resolution of 1 pc and a dark matter mass-resolution of $2001\,M_\odot$. The simulation \color{black}box is representative of a typical region of the universe (see Fig. 2 in Skinner \& Wise 2020), and \color{black}is initialized at $z=130$ using cosmological parameters consistent with the Planck 2014 constraints \citep{Planck2014}: $\Omega_\Lambda = 0.6825$, $\Omega_\mathrm{M} = 0.3175$, $\Omega_\mathrm{DM} = 0.2685$, and $h = 0.6711$. The simulation is run until $z=9.3$, when radiative transfer becomes prohibitively expensive. Its physics suite includes radiation hydrodynamics with adaptive-ray tracing \citep{wise_moray}, the nine-species (H I, H II, He I, He II, H$_2$, $e^-$, H$_2^+$, H$^-$) non-equilibrium chemistry model from \cite{Abel_1997}, radiative cooling from primordial gas and metals, momentum transfer from ionising radiation, and an $\mathrm{H}_2$-photodissociating Lyman-Werner background (LWB). \textcolor{black}{The Lyman-Werner radiation field local to each Pop III star and Pop II star cluster is modeled with an optically thin, inverse square profile with self-shielding included, and is added on top of the LWB.}

\subsection{Stars and Feedback}
\label{sec:stars_and_feedback}
 The simulation includes prescriptions for forming individual Pop III stars and Pop II star particles, described in detail in \cite{wise2012a}. Since star formation and destruction is instrumental in chemical evolution and enrichment, the details of how metals are calculated from stellar properties are restated here. When conditions for star formation are met in a cell, either a particle representing a single Pop III star \textcolor{black}{(if $[Z/H] < -5.3$)}
 \footnote{For each cell, $[Z/H]=\log_{10}\left(\frac{M_\mathrm{Z}}{M_\mathrm{H}}\right)_\mathrm{cell} - \log_{10}\left(\frac{M_\mathrm{Z}}{M_\mathrm{H}}\right)_\mathrm{Sun}$} 
 or a particle representing a cluster of Pop II stars (if $[Z/H] > -5.3$) is formed. The critical metallicity marks where dust cooling becomes eﬃcient enough to cause fragmentation at high densities (e.g., \cite{Schneider2006}.) If the particle is Pop III, the mass of the particle is sampled randomly from an IMF of the form
\begin{equation}
f(\log M)dM = M^{-1.3}\text{exp}\bigg[-\bigg(\frac{M_{\mathrm{char}}}{M}\bigg)^{1.6}\bigg]dM    
\label{eqn:mass_fn}
\end{equation}
that behaves as a Salpeter IMF at high-mass, but is exponentially suppressed below $M_{\mathrm{char}} = 20\,\,M_\odot$. \textcolor{black}{The lower and upper limits of the IMF are set to 1 $M_\odot$ and 300 $M_\odot$, respectively.} The Pop III star particle is assigned zero metallicity despite the value in the cell that formed it.  Alternatively, if $[Z/H]>-5.3$, a particle representing a coeval Pop II star cluster is formed assuming an unmodified Salpeter IMF. The particle's mass is taken to be 7\% of the cold gas within a sphere of radius $R_\mathrm{cl}$ such that mean density inside $R_\mathrm{cl}$ is $10^3$ cm$^{-3}$ (see \cite{wisecen2009} for more details). An equivalent amount of gas is removed from cells within $R_\mathrm{cl}$ of the star forming cell. The star's metallicity is initialized to the mass-weighted average of the metallicities of the surrounding cells, which can be below $[Z/H] = -5.3$ if the cell is at or slightly above threshold. As we wish the star particle to sample the massive end of the Salpeter IMF, we require that its mass exceed a minimum mass of $10^3 M_{\odot}$. If it does not, $R_\mathrm{cl}$ is increased until the condition is met. \textcolor{black}{For reference, the value of $R_\mathrm{cl}$, assuming a star particle mass of 10$^3\,\,M_\odot$ and an enclosed medium that is entirely cold, is around 2 pc.} After the Pop II star particle has lived for 4 Myr, it begins losing mass by supernovae and deposits metals continuously into the finest AMR level at every time step according to
\begin{equation}
    m_\mathrm{ej} = \frac{0.25\Delta t\times M_*}{t_0-4\text{Myr}}
    \label{eqn:p2_metal}
\end{equation}
for $t \leq t_0=20$ Myr (the lifetime of the particle), $t$ is the age of the star, and $\Delta t$ is the current time step. The ejecta has solar metallicity \textcolor{black}{$Z = 0.01295$} and is tracked in a field dedicated to metals from Pop II stars. 

Metals from Pop III stars are deposited impulsively by individual supernova events. After a Pop III star lives and radiates for its main sequence lifetime, it has different fates for different mass ranges. If $40$ M$_\odot <M_*<140$ M$_\odot$ or $M_* > 260$ M$_\odot $, the particle collapses to an inert, collisionless black hole. Otherwise, the particles explode as supernovae 
with metal ejecta masses and energies taken from \cite{nomoto2006}. If $11$ M$_\odot < M_* < 40$ M$_\odot$, the star explodes with a metal ejecta mass given by
\begin{equation}
    \frac{M_z}{M_\odot} = 0.1077+0.3383\times\left(\frac{M_*}{M_\odot} - 11\right).
    \label{eqn:type2sne}
\end{equation}
This applies to both Type II SNe ($11$ M$_\odot < M_*<20$ M$_\odot$) and hypernovae (HNe) ($20$ M$_\odot < M_* < 40$ M$_\odot$).  More massive stars in the range $140$ M$_\odot < M_* < 260$ M$_\odot$ will become PISNe \citep{heger2002} and eject metal with a mass 
\begin{equation}
    \frac{M_z}{M_\odot} = \left(\frac{13}{24}\right)\times\left(\frac{M_*}{M_\odot}-20\right)
    \label{eqn:pisn}
\end{equation}
at the end of their lifetime.

The blast wave is modeled by injecting the explosion energy and ejecta mass into a sphere of 10 pc, smoothed at its surface to improve numerical stability. \textcolor{black}{Typically, an explosion occuring in a medium with density $\sim1000$ g/cm$^3$ would would need to be resolved on scales of $<1$ pc in order to resolve the Sedov-Taylor phase \citep{KimOstriker2015}, but because the density of the medium surrounding the Pop III star is reduced by photoevaporation during the star's lifetime \citep{Whalen2004}, the medium in which explosion energy is deposited has a larger cooling radius, and the choice of 10 pc is thus sufficient to resolve the Sedov-Taylor phase in this case.} After its destruction, the star is converted to a collisionless particle \textcolor{black}{with a mass of 10$^{-20}$ times the mass of the original Pop III star. This renders the particle mass negligible while still containing information about the progenitor star and allowing one to approximately track the center of mass position of the supernova remnants.} The metal contribution from the explosion is logged into a separate metallicity field for Pop III stellar ejecta. Explosion energies are assigned as follows: Type II SNe have $E_{51}=1$; HNe have $10 \leq E_{51} \leq 30$ depending on their mass; and PISNe have $6.3 \leq E_{51} \leq 90$, according to Eq. 3 of \cite{wise2012a}. Here $E_{51}$ is the explosion energy in units of $10^{51}$ erg. For our choice of Pop III IMF, the relative occurence of supernovae of different types are 38\% Type II SNe, 54\% HNe, and 8\% PISNe. We chose a Pop III characteristic mass so that HNe would dominate the chemical enrichment process, as the chemical abundance of \textcolor{black} {Mn, Co, Ni, Zn, Ca, and Cr relative to Fe} of Extreme Metal Poor stars are better fit by hypernova models \citep{nomoto2006}. 
\label{sec:RS}

\section{Enrichment of Halos}
\label{sec:enrichment}


\begin{figure*}[t]
    \centering
    \includegraphics[width=0.8\textwidth]{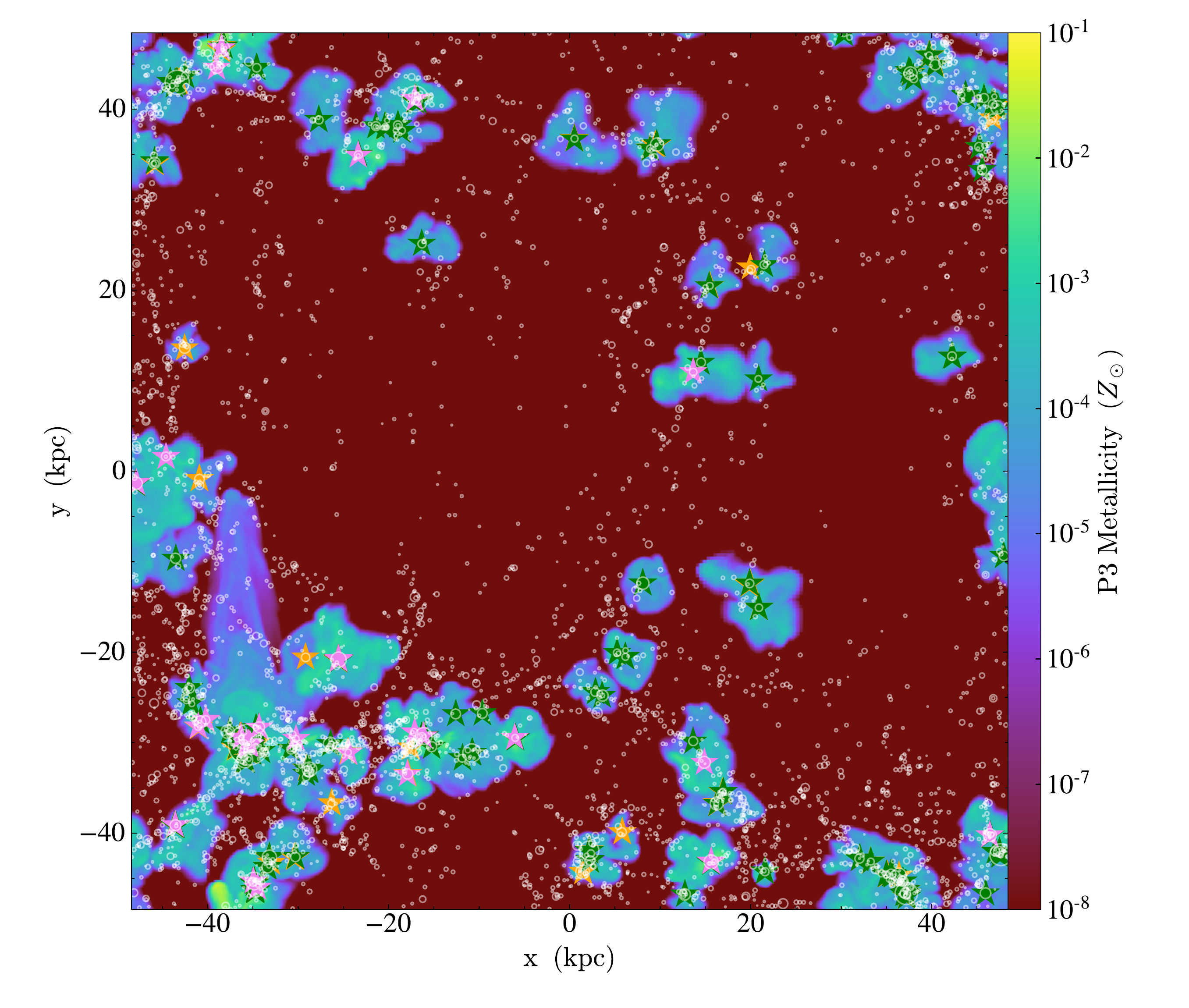}
    \caption{Pop III metallicity projection of the simulation at $z=9.3$. The white circles outline the halos with virial mass greater than $10^{5.3}\,M_\odot$ identified with ROCKSTAR. The radius of each circle corresponds to the virial radius of the halo it represents. The stars label the Pop III SN remnant particles in the simulation (orange: Type II SNe, green: HNe, violet: PISNe). The distance scales here are in proper coordinates.}
    \label{fig:projection_annotated_DD1030}
\end{figure*}

\begin{figure}[t]
    \centering
    \includegraphics[width=0.47\textwidth]{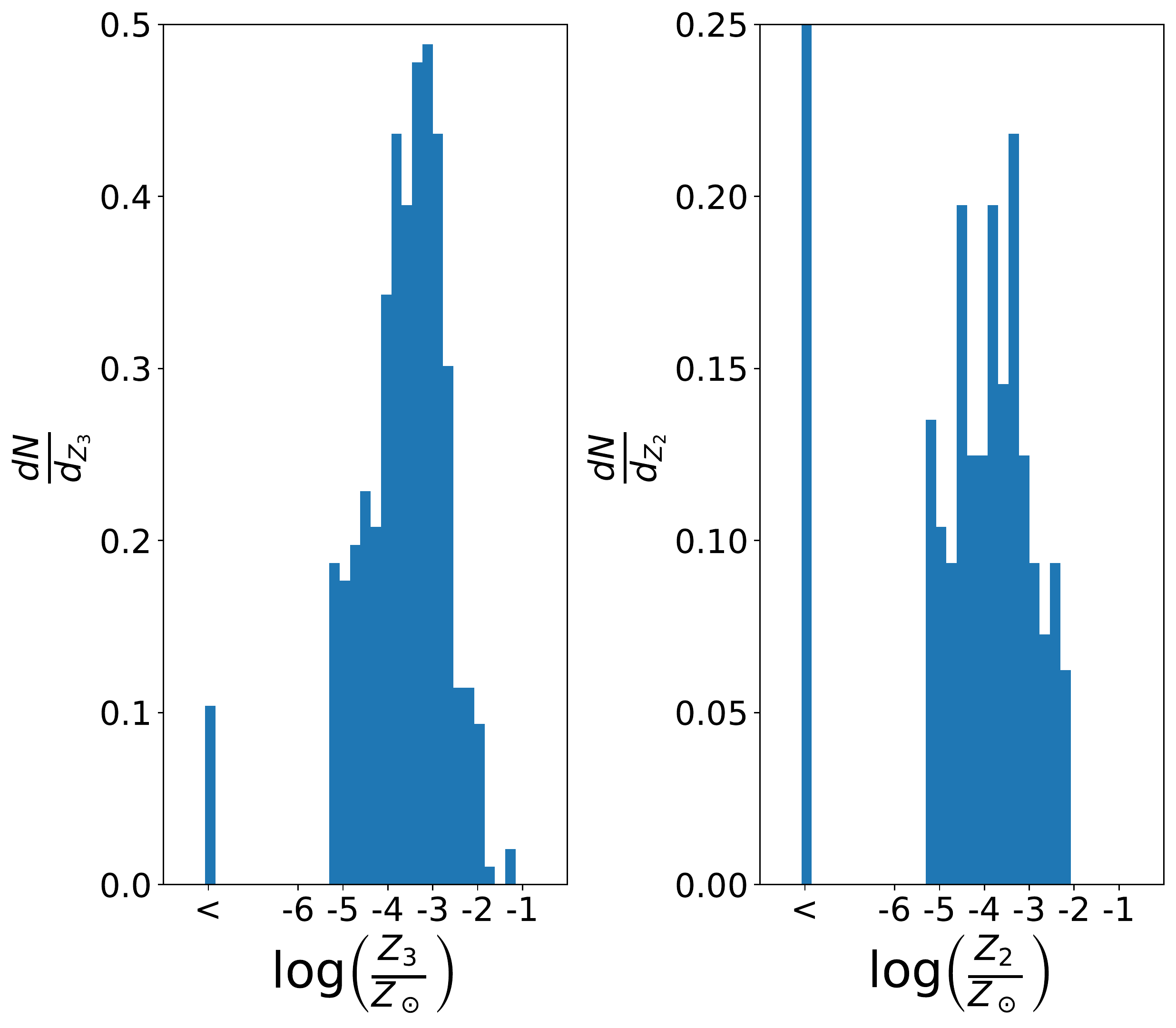}
    \caption{\textcolor{black}{Mass-weighted Pop III and Pop II metallicity distributions over enriched halos at $z=9.3$. The bin labeled `$<$' contains all of the halos with metallicity below $Z_\mathrm{crit}=10^{-5.3}\,Z_\odot$ for the respective metallicity type. The median metallicities for the respective distributions above $Z_\mathrm{crit}$ are $[\langle Z_3 \rangle/H]=-3.44$ and $[\langle Z_2 \rangle /H]=-3.62$. For the sample of 417 enriched halos, 84 are dominated by Pop II metals.}}
    \label{fig:mdf}
\end{figure}

\begin{figure*}[!ht]
    \centering
    \includegraphics[width=0.94\textwidth,height=0.94\textheight]{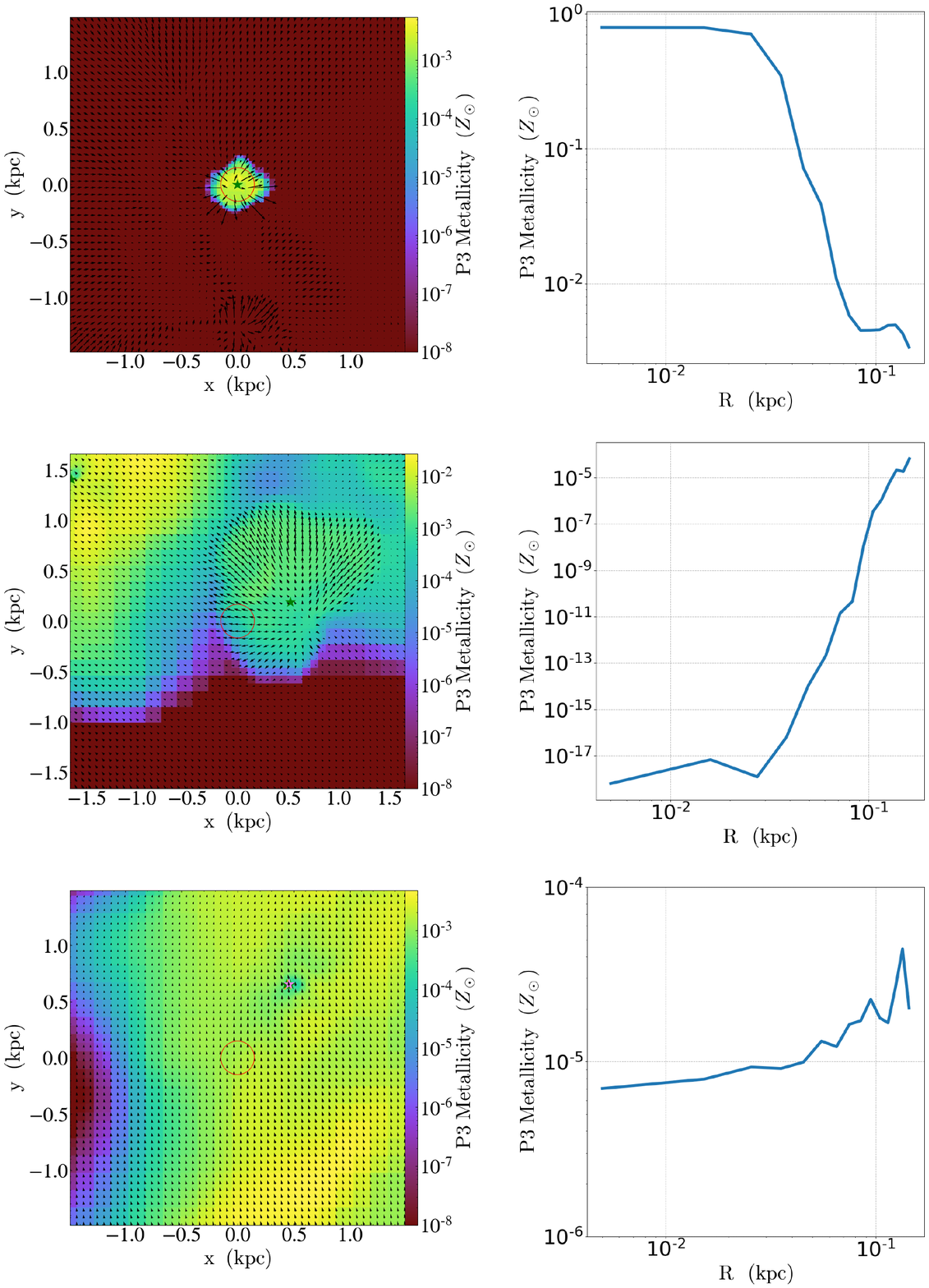}
    \caption{Examples of internally enriched (\textbf{top}), externally enriched (\textbf{middle}), and pre-enriched (\textbf{bottom}) halos just after the total mass-weighted average metallicity passes $Z_\mathrm{crit}$. \textbf{Left:} Pop III metallicity \textcolor{black}{projection of a cubic subvolume} centered on the halo indicated by the red circle, which has radius equal to the halo's virial radius. Arrows indicate the velocity fields, and the stars show the position of Pop III remnant particles (orange: Type II SNe, green: HNe, violet: PISNe). \textbf{Right:} Pop III metallicity profile for the halo binned by proper distance from the center. The ordinate axis shows the mass-weighted average metallicity within each bin. \textcolor{black}{All distance scales are in proper coordinates.}}
    \label{fig:exemplary_figure}
     
\end{figure*}


For the following analysis, we define the variable,
\begin{equation}
    f_3\equiv\log_{10}{\left(\frac{M_{\zpiii,\mathrm{p}}}{M_{\zpiii,\mathrm{a}}}\right)},
    \label{eqn:f3}
\end{equation}
to describe the enrichment of halos.  Here, $M_{Z_3,\mathrm{p}}$ is the maximum mass of Pop III metals that could have originated from Pop III stars within the virial radius of each halo. $M_{Z_3,\mathrm{p}}$ is calculated using Equations \ref{eqn:type2sne} and \ref{eqn:pisn}, given the Pop III remnant particles inside the virial radius at the final redshift. $M_{Z_3,\mathrm{a}}$ is the Pop III metal mass within the virial radius that is directly measured from the simulation. We label a halo as ``enriched" if it satisfies $\langle\zpii\rangle + \langle\zpiii\rangle \equiv \langle Z_{\mathrm{total}} \rangle > 10^{-5.3}\,Z_\odot$, where $\langle\zpii\rangle$ and $\langle\zpiii\rangle$ are the mass-averaged Pop II and Pop III metallicities inside the virial radius. Otherwise, the halo is labeled as ``pristine." Using these definitions, $f_3 < 0$ corresponds to halos that contain more metals than could have come from internal sources, meaning that external enrichment must have occured; $f_3 = -\infty$ corresponds to halos that are enriched exclusively by external sources, and $f_3 > 0$ corresponds to halos that are primarily enriched by internal sources. A third case of enrichment exists for halos that form within a region of the IGM that has already been enriched. We label this subset of halos as pre-enriched, and discuss their prevalence in our sample in Section \ref{sec:pre_enriched}.

\color{black}
An issue with this measure exists for the case where a halo undergoes a mix of internal and external enrichment, but is dominated by external enrichment in such a way that the total amount of metals confined within the halo is still less than the predicted value based on internal events $M_{Z_3,\mathrm{p}}$. This would result in a value of $f_3>0$ for a halo that is still primarily externally enriched. We argue that this is very a very unlikely occurance because (1) very few halos actually experience internal supernova events and are still dominated by external sources (this is explored in Section \ref{sec:ext_enrichment}), and (2) this also requires most of the ejecta from the internal event to make it outside of the virial radius, which, while not necessarily uncommon given the assumed Pop III characteristic mass, further limits the likelyhood of an externally enriched halo being mislabeled in this way.

\color{black}
An example halo for each type of enrichment is shown in Figure \ref{fig:exemplary_figure}. The left column shows a projection of each type of enriched halo just after the total mass-weighted average metallicity passes $Z_\mathrm{crit}$ while the right column shows the Pop III metallicity profile of each halo at the same snapshot as the corresponding projection. For the internally enriched case, a Pop III hypernova event occurs near the center of the selected halo. The corresponding Pop III metallicity profile shows large, near-solar metallicities towards the center, with decreasing values going out to the virial radius. The external enrichment example shows the opposite behavior. A hypernova event occurs outside the virial radius, and, because there is turbulent mixing from the outside-in, the metallicity profile shows the highest values towards the edges of the halo, with decreasing values going towards the center, which is composed of nearly pristine gas. The example for pre-enrichment shows a halo that forms inside a medium that was already enriched by a nearby PISN event. The metal ejecta has had time to mix with the surrounding medium, so the resultant Pop III metallicity profile for the pre-enriched halo is nearly flat, increasing outwards by less than an order of magnitude.  
\color{black}
\subsection{Externally Enriched Halos}
\label{sec:ext_enrichment}

\begin{figure}[t]
    \centering
    \includegraphics[width=0.47\textwidth]{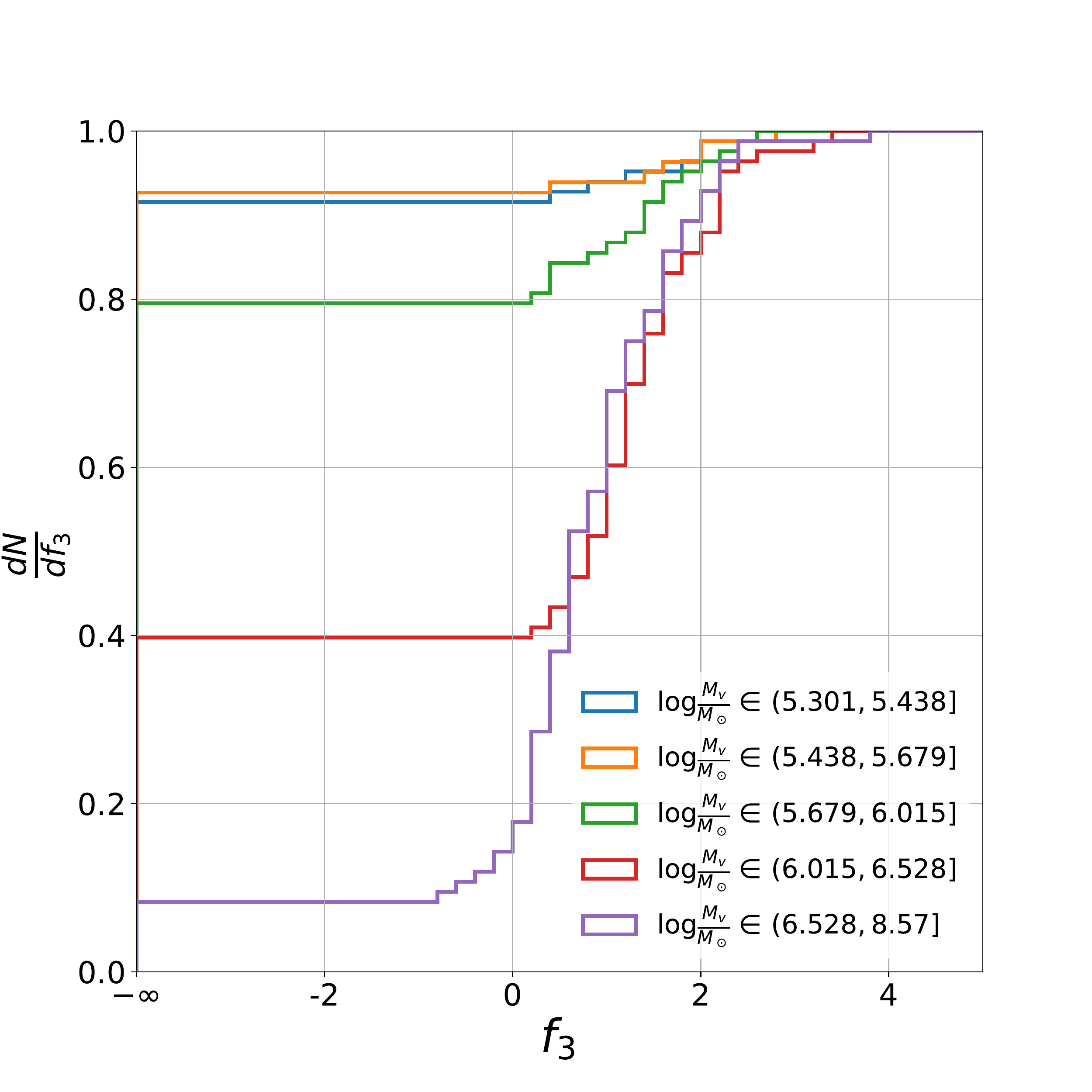}
    \caption{Cumulative distribution function of $f_3\equiv\logR$ for the 417 enriched halos in the simulation at the final redshift $z=9.3$. The halos are binned according the virial mass such that each bin contains approximately the same number. The majority of externally enriched halos ($f_3 < 0$) are enriched purely from outside sources ($f_3 = -\infty$). Additionally, most of the externally enriched halos have $M_\mathrm{vir} < 10^{6} M_\odot$.}
    \label{fig:f3_plot}
\end{figure}

The first question concerning external enrichment is simple: does external enrichment happen in an appreciable number of halos? To answer this question, halos are identified using ROCKSTAR \citep{0004-637X-762-2-109}, and the value of $f_3$ is calculated for each. We find that the number of both internally and externally enriched halos increases over time, and as redshift decreases, external enrichment becomes the most common enrichment vector.  At the final output, $z=9.3$, the simulation hosts 1864 halos, \textcolor{black}{including subhalos}, with virial mass above our 100-particle resolution limit of $10^{5.3}\,M_\odot$ and \textcolor{black}{417 halos that are enriched to $\langle Z_{\mathrm{total}} \rangle > 10^{-5.3}\,Z_\odot$. Of the enriched halos, 264 (63.3 \%) are found to have $f_3 < 0$}.

Figure \ref{fig:projection_annotated_DD1030} shows a Pop III metallicity projection of the simulation at $z=9.3$ indicating the location and virial radius of each halo and the location and type of all SN remnants. The regions with metallicities in excess of $Z_\mathrm{crit}=10^{-5.3} Z_\odot$ tend to show more halo clustering. The halo-averaged Pop III and Pop II metallicity distribution functions are shown in Figure \ref{fig:mdf}. The regions of high spatial density are where halos are more likely to be enriched externally. A calculation of the mean halo-to-halo nearest-neighbor distance gives a value of \color{black}0.66 \color{black}proper kpc for externally enriched halos, whereas the mean nearest neighbor distance across all halos is around \color{black}1.6 \color{black}proper kpc. There is one region near $(x,y)=(-35\,\mathrm{kpc},-30\,\mathrm{kpc})$ of Fig. \ref{fig:projection_annotated_DD1030} that contains a particularly large volume of enriched gas that is the result of mixing between ejecta of many SN explosions. This region hosts the most massive halo in the simulation.

\color{black}
Figure \ref{fig:f3_plot} shows the $f_3$ cumulative probability distribution of enriched halos at $z = 9.3$ separated into five mass bins each containing the same number of halos between $10^{5.3}\,M_\odot$ and $10^{8.6}\,M_\odot$. The fraction of halos that undergo pure external enrichment generally increases with decreasing halo mass. Below $10^6\,M_\odot$, the majority of halos are found to be enriched purely externally, while most of the halos above $10^6\,M_\odot$ are enriched internally. There is a small minority of halos in the bins above $10^6\,M_\odot$ that experience both internal and external enrichment by our measure, while the enrichment pathway for the remaining halos is either pure internal or pure external. In some cases (e.g., lower left-hand corner of Fig. \ref{fig:projection_annotated_DD1030}), there is one high-mass halo that injects large amounts of metals into its surroundings and draws multiple low-mass satellites into the enriched region, turning them into externally enriched halos.
\color{black}


\begin{figure}[t]
    \centering
    \includegraphics[width=0.47\textwidth]{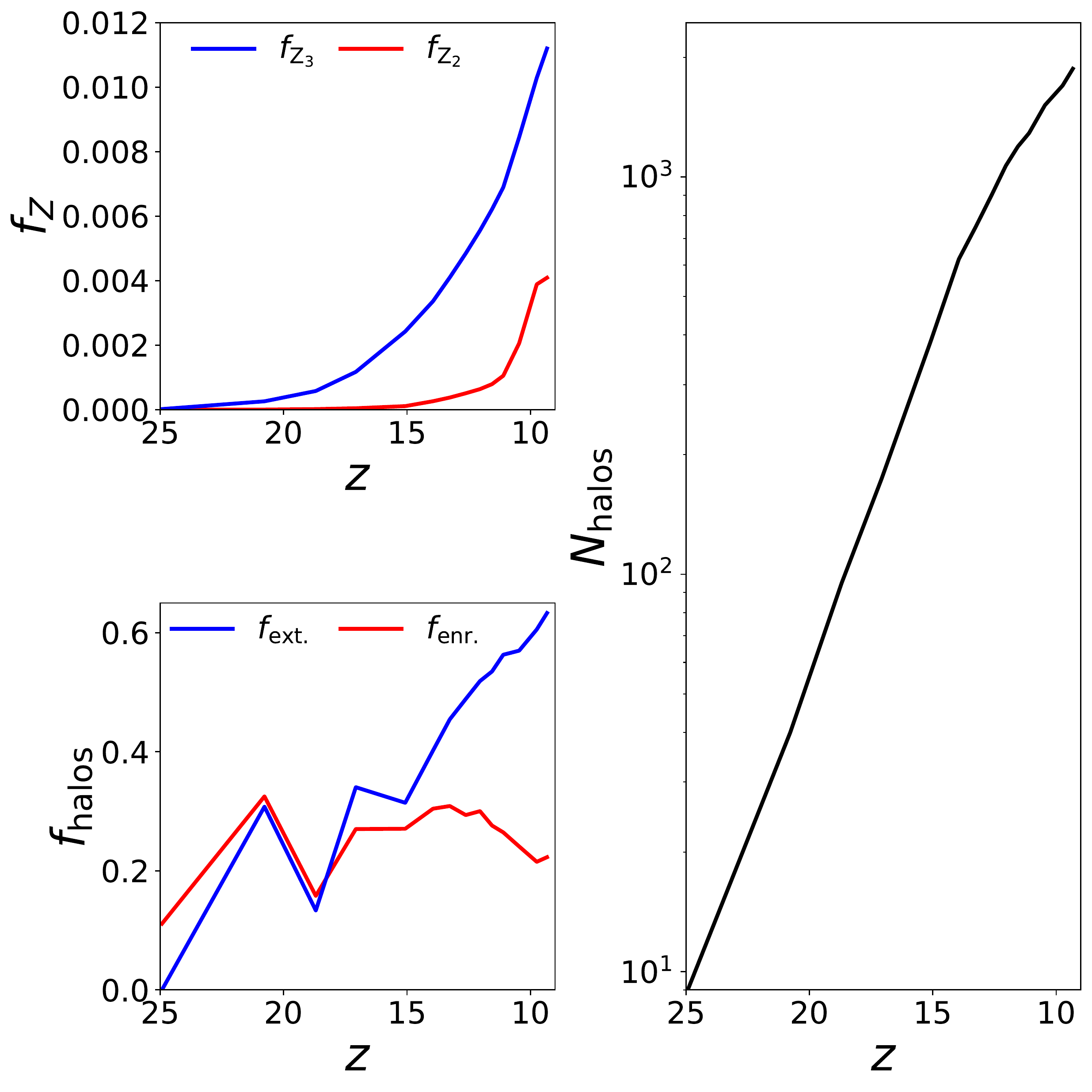}
    \caption{Number fractions of various quantities as a function of redshift. \textbf{Top left:} volume filling fraction for gas with $Z_3 > 10^{-5.3} Z_{\odot}$ (blue) and $Z_2 > 10^{-5.3} Z_{\odot}$ (red). \textbf{Bottom left:} The red line shows the fraction of all halos above our mass cutoff that are enriched. The blue line shows the fraction of enriched halos that are found to be externally enriched. \textbf{Right}: Number of halos above the minimum mass cut as a function of redshift.}
    \label{fig:fractions}
     
\end{figure}

Figure \ref{fig:fractions} shows the evolution of various number fractions over time. While the fraction of all halos that are enriched above $Z_3=10^{-5.3}\,Z_\odot$ stays roughly constant, the fraction of those that are externally enriched increases over time. The externally enriched halo fraction follows the also-increasing volume fraction of enriched gas in the simulation, suggesting that the growing body of metal ejecta overtakes the halos over time or that the number of halos forming in the enriched IGM is increasing.


\color{black} 
\subsection{Pre-enriched vs. externally enriched?}
\label{sec:pre_enriched}
So far, we have assumed that every enriched halo that is not internally enriched is externally enriched by a nearby supernova remnant. However, halos forming in an enriched region of the IGM will accrete gas from the environment during virialization and become enriched that way. We call this kind of halo pre-enriched. As the volume fraction of enriched gas increases, we would expect more pre-enriched halos to form. Here we analyze the occurrence of pre-enrichment in our simulation. We flag halos as pre-enriched if their mass-weighted metallicity is above $10^{-5.3} Z_{\odot}$ in the first data output in which they appear; i.e., become resolved in our simulation. The time interval between the data outputs \textcolor{black}{used here is 2-4 Myr, which is} less than the lifetime of a typical Pop III star, so it is unlikely we confuse an externally enriched halo with a pre-enriched halo. 

\begin{figure}[t]
    \centering
    \includegraphics[width=0.42\textwidth]{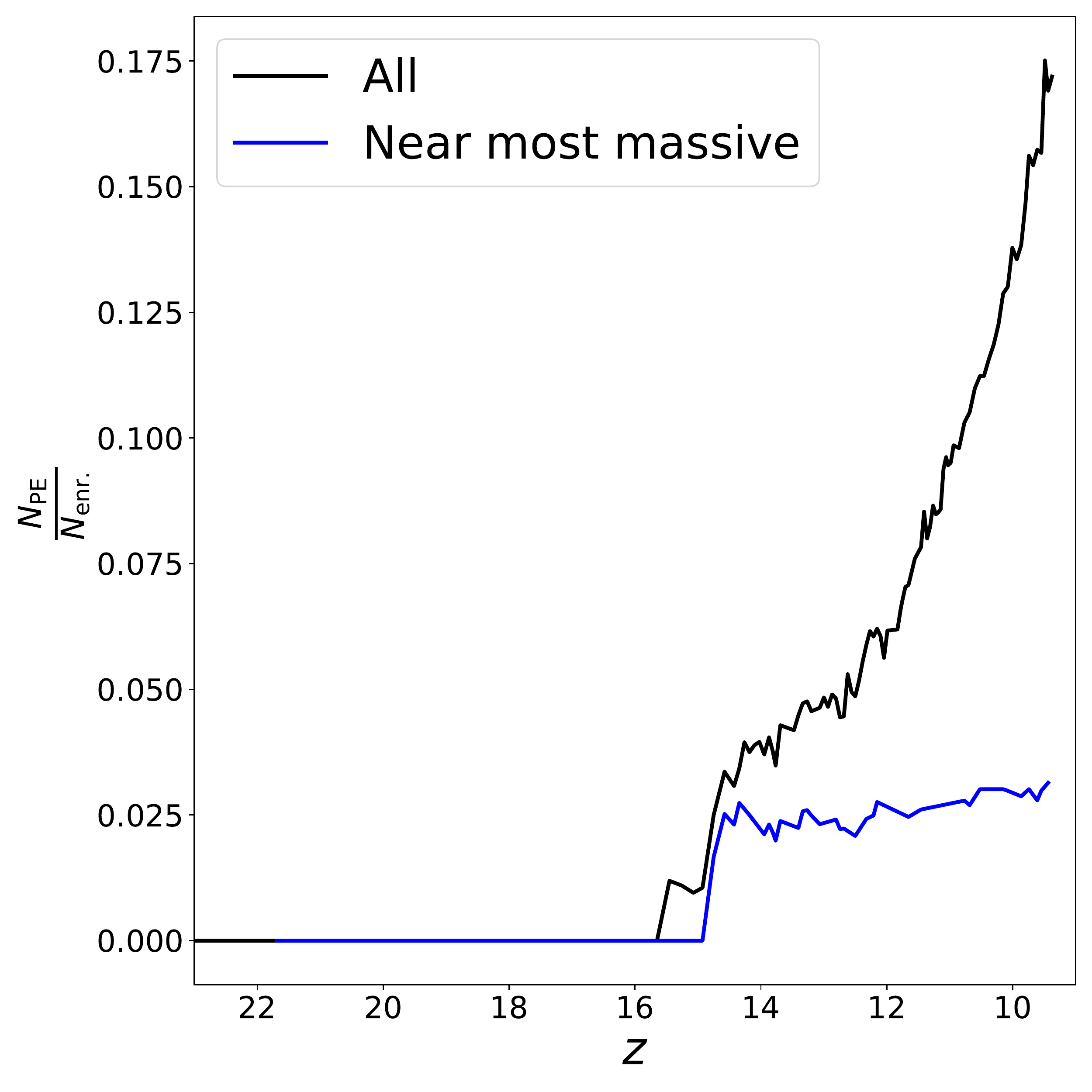}
    \caption{Formation of pre-enriched halos. Fraction of all enriched halos that were enriched when they were initially resolved versus redshift. Black line is for all halos in the volume; blue line is for halos within 10 proper kpc of the most massive halo's most massive progenitor. Pre-enrichment is a subdominant channel relative to internal and external enrichment of minihalos at all redshifts simulated, growing from a few percent at $z\sim 15$ to \color{black}$17.5 \%$ \color{black}at $z=9.3$.}
    \label{fig:pre-enriched}
\end{figure}

Fig. \ref{fig:pre-enriched} plots the ratio of the number of pre-enriched halos to all enriched halos versus redshift. The black line is for the entire volume, while the blue line is only for halos within 10 proper kpc of the most massive halo. We see that pre-enriched halos form predominantly near the most massive galaxy for $z>12$, but form throughout the volume in increasing numbers at lower redshift. At $z=9.3$, 17.5\% of all enriched halos formed pre-enriched, implying that external enrichment by nearby supernova remnants remains the primary channel for halos that are not internally enriched.

\subsection{Contribution of Different SN Types}


\begin{figure*}[t]%
    \centering
    \subfigure{%
        \includegraphics[width=0.47\textwidth]{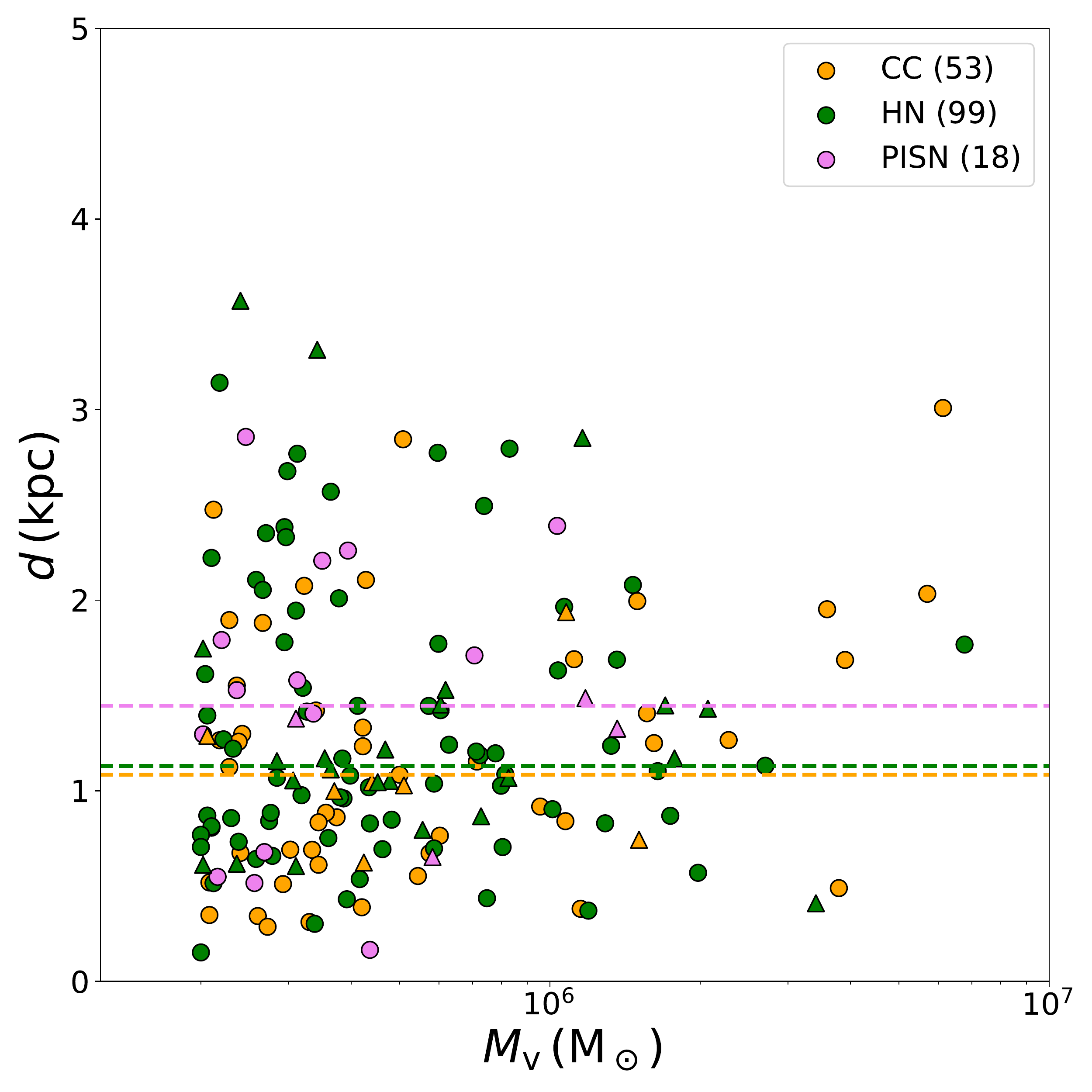}}%
    \qquad
        \subfigure{%
        \includegraphics[width=0.47\textwidth]{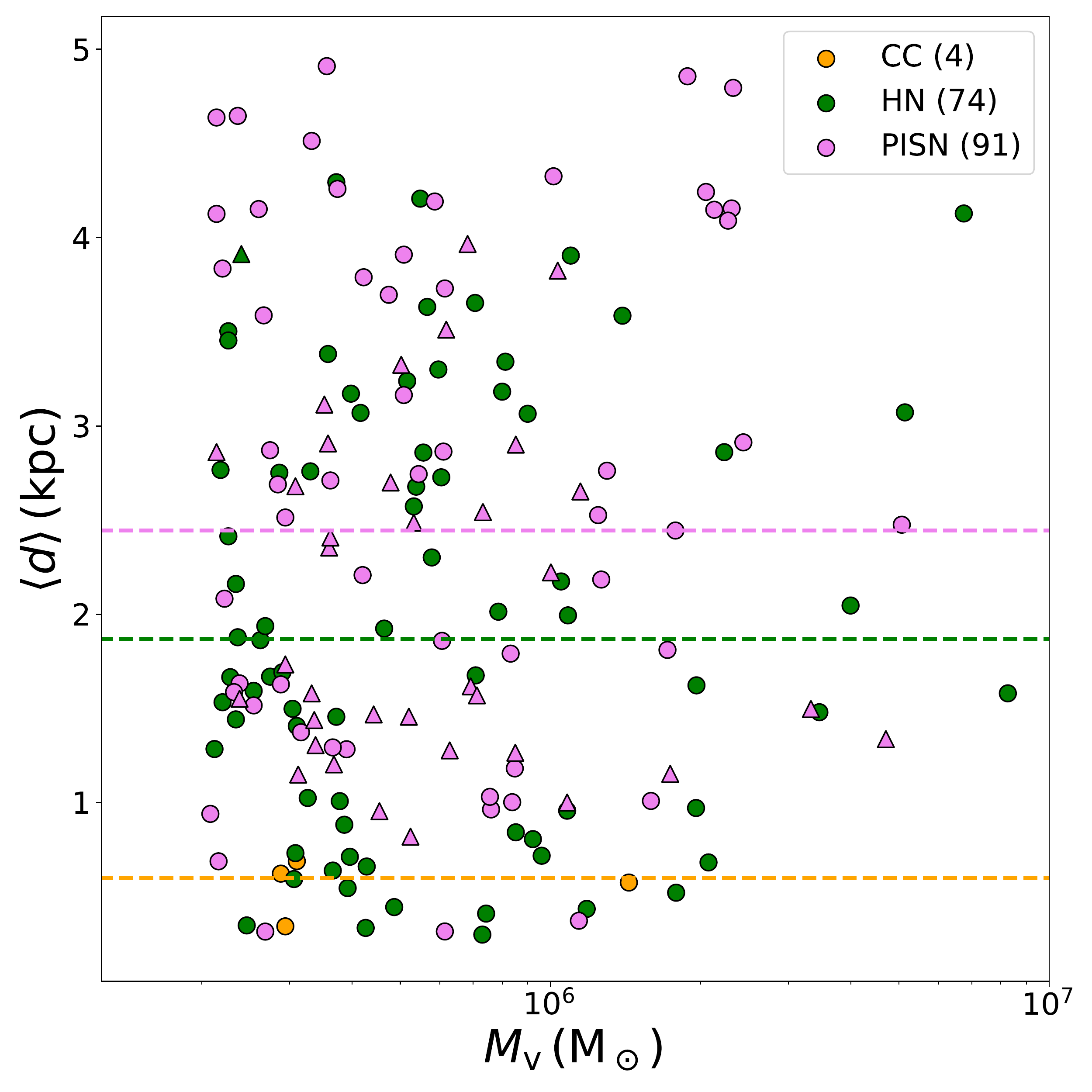}}
        
    \caption{\textbf{Left:} Proper distance from each externally enriched halo, identified at the final output, to the Pop III remnant particle at the time it first reaches the metallicity threshold of $10^{-5.3}\,\,Z_\odot$ vs. virial mass at that time. \textbf{Right:} Mean distance to the Pop III remnant particles of the dominant enriching type within 2 kpc at the final output vs. virial mass at the final output. Halos that do not have any Pop III remnant particles within a 2 proper kpc radius are not plotted. Colors represent the type of Pop III remnant. The horizontal dashed lines represent median distances within the respective bins. Triangles represent halos that are within 10 proper kpc of the most massive halo at the final output. The parentheticals in the legends show the number of halos within each SN-type bin, without distinguishing between halos that are close to the most massive halo and those that are not. \textcolor{black}{Both panels exclude pre-enriched halos.}}
\label{fig:rem_dists}
\end{figure*}

In order to characterize the contribution of each type of supernova in the simulation to the external enrichment process, we perform the following calculations. First, we identify the type of the nearest Pop III remnant to each externally enriched halo at the output when the halo first crosses the metal enrichment threshold of $10^{-5.3}\,Z_\odot$ and calculate the proper distance from the halo's center to the remnant at that output. We then calculate the total ejecta of all Pop III remnants within a 2 proper kpc radius of each externally enriched halo at the final output using Equations \ref{eqn:type2sne} and \ref{eqn:pisn}, and consider the type that has produced largest summed Pop III metal mass to be the dominant enricher of that halo. The mean distance from the externally enriched halo to all Pop III remnants of the dominant enriching type is then recorded. The results of these calculations are shown in Figure \ref{fig:rem_dists}. The left panel shows the results of the first calculation, while the right panel shows the results of the second.

As shown in the right panel of Figure \ref{fig:rem_dists}, \textcolor{black}{54\% of externally enriched halos are enriched primarily by PISNe by the last output, while approximately 44\% and the remaining 2\% are enriched primarily by HNe and Type II SNe, respectively.} Taking into account the relative abundances of SN types set by the $20\,M_\odot$ Pop III characteristic stellar mass (54$\%$ are HNe, 38$\%$ are Type II SNe, and 8$\%$ are PISNe), \textcolor{black}{it is surprising that PISNe are able to surpass HNe as the dominant enricher of the largest fraction of halos. While the statistics on this are poor, it suggests that PISNe enrich more halos per event on average, which is understandable because PISNe are $3-10 \times$ times more energetic than HNe and produce $\sim10$ times more metals on average for our model.}

The most massive halo in the simulation is a significant source of metal enrichment, enclosing 62 HN, 45 Type II SN, and 7 PISN Pop III remnant particles by the last output. This is the most active region in the simulation, and the metallicity field surrounding the most massive halo is the result of mixing between many supernova remnants of different types. Halos that form in this environment could initially be labeled as externally enriched by our measure, and the type of Pop III supernova that contibutes the most to their enrichment is less clear. Because of this, the externally enriched halos that are within 10 proper kpc of the most massive halo by the final output are also identified in Figure \ref{fig:rem_dists}. \textcolor{black}{Of the 169 halos plotted in the right panel, 36 are within 10 proper kpc of the most massive halo}. As seen in the right panel, the majority of halos nearby the most massive halo are identified as having been enriched primarily by PISNe by our measure. This makes sense because of the large concentration of PISN remnant particles in that particular region.

The median lines in Figure \ref{fig:rem_dists} give an indication of typical distances by which enrichment from each type of supernova can occur. In both panels, there is a clear stratification between distances to Type II SNe, HNe, and PISNe (listed in order of increasing distance). In the left panel, the median distance to the nearest Pop III remnant particle at the time of enrichment is \textcolor{black}{1.08 proper kpc for Type II SNe, 1.13 proper kpc for HNe, and 1.45 proper kpc for PISNe. In the right panel, the median distances are 0.599 proper kpc for Type II SNe, 1.87 proper kpc for HNe, and 2.44 proper kpc for PISNe}. It should be noted that the Type II SNe bin in the right panel of Figure \ref{fig:rem_dists} only has four data points, so the median is less reliable for that one bin. The ordering in median distance to each type makes sense on energetic grounds, as PISNe are more energetic than HNe, which are more energetic than Type II SNe (see Sec. 2).
\color{black}

In order to further verify that the distances found in Figure \ref{fig:rem_dists} are reasonable, the following estimate is performed. The typical enriching radius of each type of supernova is calculated by considering the average volume that each type enriches to \textcolor{black}{$\zpiii > Z_\mathrm{crit}$}. This is calculated as follows:
        $$\langle V_{\mathrm{enr}} \rangle_{\mathrm{type}}=f_{\mathrm{ej}}\left(\frac{V_{\mathrm{enr}}}{N_{\mathrm{type}}}\right)$$
        
        $$\langle r_{\mathrm{enr}} \rangle_{\mathrm{type}}=\left[\frac{3}{4\pi}\langle V_{\mathrm{enr}} \rangle_{\mathrm{type}}\right]^{1/3}$$

\noindent Here, $f_{\mathrm{ej}}$ is the fraction of the total mass of ejected metals from each type of SN by $z=9.3$, $N_\mathrm{type}$ is the number of each type that has occured, and $V_\mathrm{enr}$ is the total volume of enriched gas at $z=9.3$. By the final output, $93\%$ of explosions that have occured are Type II core-collapse SNe or HNe, while the remaining $7\%$ are PISNe. By $z=9.3$, there are 169 Type II SNe, 240 HNe, and 32 PISNe that have occured. Interestingly, the few PISN explosions that occur contribute more metal ejecta than both of the other SN types combined. Using the ejecta Equations \ref{eqn:type2sne} and \ref{eqn:pisn}, these contribute $212.2\,M_\odot$, $1{,}494\,M_\odot$, and $2{,}753 \,M_\odot$ of Pop III metal ejecta in each respective bin. Applying the formulae above, the calculation yields values of approximately \textcolor{black}{0.97 proper kpc, 1.5 proper kpc, and 3.6 proper kpc} for the average enriching radius of Type II SNe, HNe, and PISNe respectively. These radii are in rough agreement with those of the supernova remnants simulated in \cite{Whalen2008}, the visual presented in Figure \ref{fig:projection_annotated_DD1030}, and the median distances in Figure \ref{fig:rem_dists}. It should be noted that this calculation does not take into account the age of the remnants, as an older remnant has had more time to expand. The calculation also does not account for overlapping SN remnants, gas collapse, or mixing as a result of halo mergers. The radii derived above therefore must be viewed as rough upper limits. A more detailed calculation would use the individual mass of each SN progenitor, rather than assigning the average bubble size to all SNe that are labeled as a given type, while adjusting $V_\mathrm{enr}$ appropriately for gas dynamics. It should also be noted that these results are highly dependent on the IMF chosen for Pop III star formation.

\section{Star Formation}
\label{sec:star_formation}

\subsection{The First Pop II Stars}
\begin{figure}[t]%
    \centering
    \includegraphics[width=0.47\textwidth]{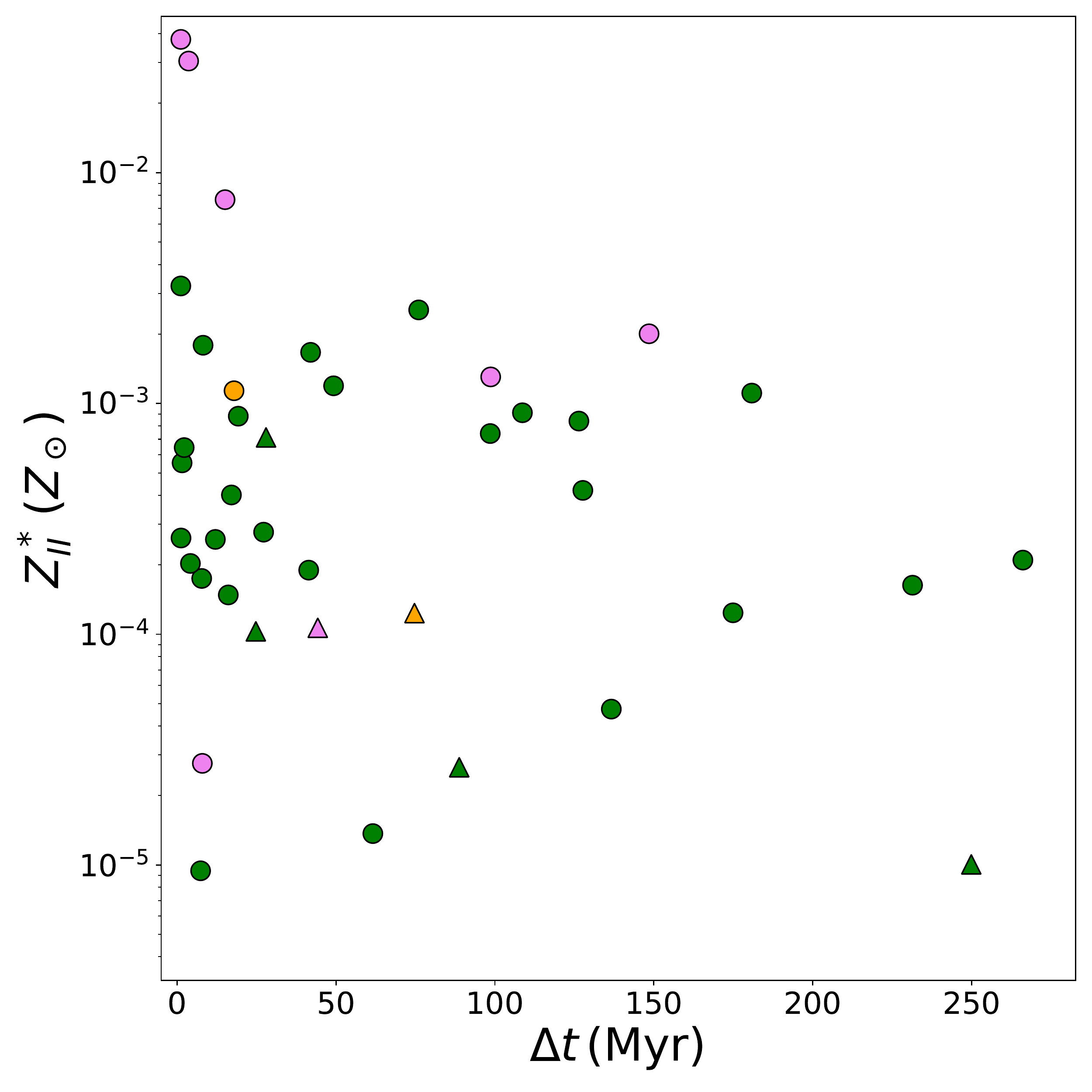}
    \caption{Metallicity and time delay of the first Pop II star particle to form after the first Pop III supernova for each star-forming halo. The symbol shapes indicate the enrichment process leading to the formation of the first Pop II star particle (circles: internal enrichment, triangles: pure external enrichment) and the symbol colors represent the supernova type of the first Pop III remnant particle that ends up in the final halo (orange: Type II SNe, green: HNe, violet: PISNe). The halos whose first Pop II stars form via pure external enrichment form in halos with $M_v<10^6\,M_\odot$, with metallicities below $10^{-3}\,Z_\odot$. Points corresponding to Pop II stars that form in halos below the minimum mass cut are not shown. The median time delay for all halos plotted, only internally enriched halos, and only externally enriched halos is 28, 19.3, and 59.5 Myr, respectively.}  
    \label{fig:p3p2_scatter}
\end{figure}

Recall that in this simulation Pop II star particles represent coeval star clusters formed out of gas that has been enriched by Pop III supernovae and/or prior generations of Pop II star formation. One interesting question is how sensitive are the characteristics of the first Pop II stars to form to the type of the first Pop III supernova to occur in the host halo's history, as well as to the enrichment pathway leading to their formation. To shed light on this matter, for each halo, we compare the creation times of all Pop II particles and Pop III remnant particles within the halo's virial radius at the final output. We then select the earliest particle of each type for further examination. The explosion type of the first Pop III supernova is logged, and an additional check is performed to determine if the first Pop II star particle formed as a result of external enrichment. Here, a halo is considered externally enriched if it forms a Pop II star particle without a Pop III remnant particle inside the virial radius. The metallicity requirement from Sec. \ref{sec:enrichment} has been dropped because we only consider halos that form Pop II stars, and the formation of a Pop II star particle is a direct indication that the halo has been enriched by some process. If the first Pop II star particle forms inside a subhalo of a halo that contains Pop III remnant particles, then the Pop II particle is determined to have formed through external enrichment as long as none of the Pop III remnant particles are inside the subhalo at the time of the Pop II particle's formation. Some \color{black}17\% \color{black}of all externally enriched halos that are not pre-enriched are subhalos of larger halos.  

Figure \ref{fig:p3p2_scatter} shows the type of the first Pop III supernova, the metallicity of the first Pop II star particle, and the time delay between the first Pop III supernova and the formation of the first Pop II particle for each star-forming halo. Note that Pop II star particles exist with metallicities below $Z_{\mathrm{crit}}=10^{-5.3} Z_{\odot}$ because they are assigned the mass-weighted average metallicity of their birth cloud, which can be less than $Z_{\mathrm{crit}}$. Because of our choice of 20 $M_\odot$ for the Pop III characteristic mass, 78\% of the star-forming halos are originally seeded by hypernovae, which explains their prevalence. PISNe are much rarer, again because our choice for the primordial IMF. The Pop II particles that form following a PISN tend to have higher metallicity, with an average \color{black}in $\log{\left(Z_\mathrm{II}^*\right/Z_\odot)}$ of $10^{-2.7}\,Z_\odot$ compared to the average for particles forming after HNe and Type II SNe of $10^{-3.5}\,Z_\odot$\color{black}. The first Pop II star particles with the highest metallicities form $<5$ Myr after a PISN. However, there are only 7 star-forming halos that are seeded by PISNe, and for 2 of these halos, the first Pop II star particles have metallicities below $10^{-5.3}\,Z_\odot$. Further study on this topic would require a larger sample size to draw reliable conclusions.  Also shown in Fig. \ref{fig:p3p2_scatter} are the halos that form their first stars through external enrichment. Of the 41 star forming halos in this sample, 5 formed their first Pop II stars through external enrichment. All of the particles that formed through external enrichment, including those that form following PISN explosions, have metallicity below $10^{-3}\,Z_\odot$.

\begin{figure}[t]
    \centering
    \includegraphics[width=0.47\textwidth]{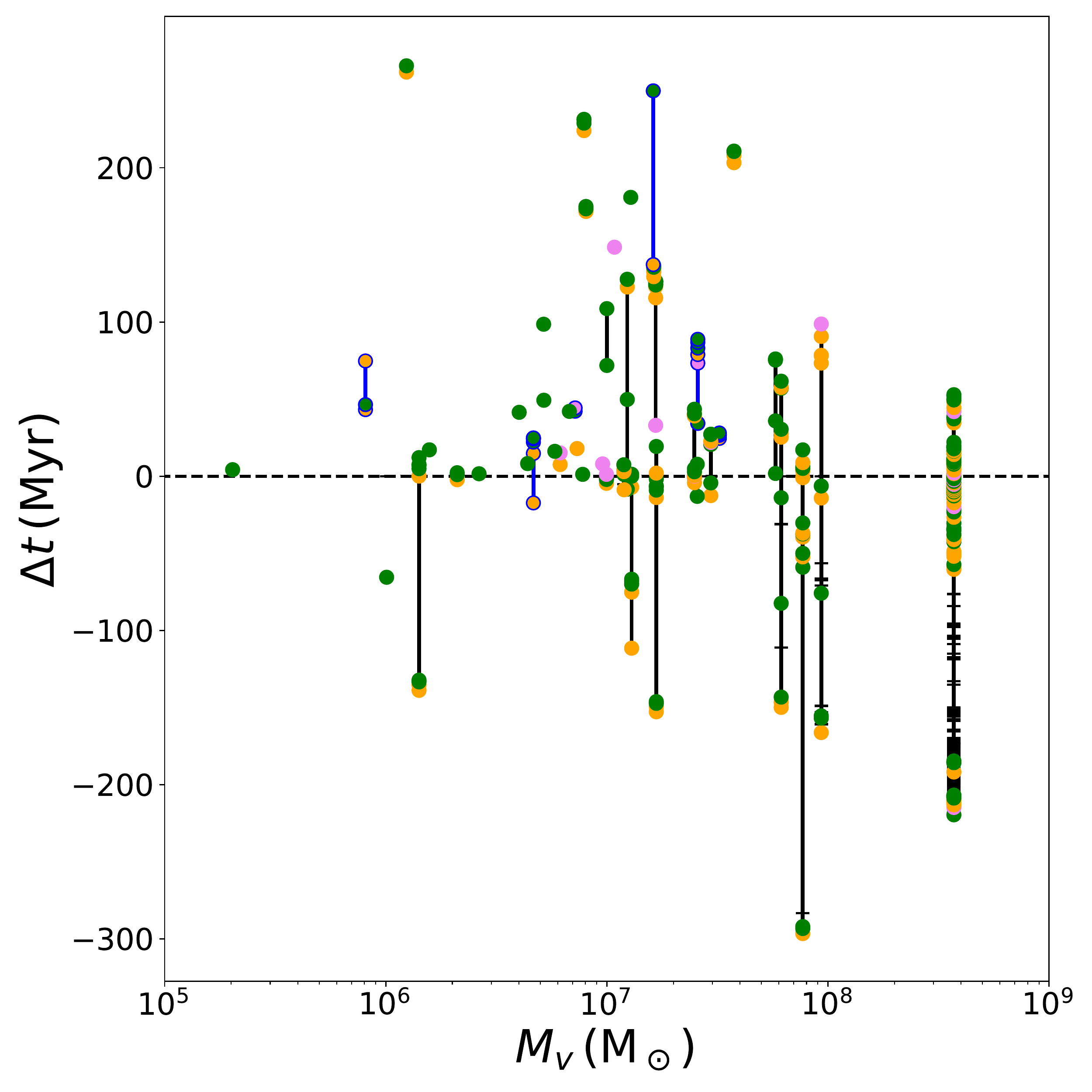}

    \caption{History of all Pop III supernovae in each Pop II star-forming halo's history, relative to the formation time of the first Pop II star particle ($\Delta t \equiv t^{\mathrm{first}}_{II} - t_{III}$). Positive values correspond to events happening before the first Pop II star particle's formation while negative values correspond to events happening after. Points connected by a vertical line are associated with the same final halo whose virial mass is indicated by the abscissa. The color of the points correspond to the type of each supernova (orange: Type II SNe, green: HNe, violet: PISNe). Horizontal dashes on the vertical lines show the times of further Pop II star formation during the Pop III-Pop II overlap phase. The points outlined in blue and connected with blue lines correspond to halos that form their first stars through external enrichment.}
    \label{fig:p3p2_overlap}
\end{figure}

 The Pop III and Pop II epochs of star formation are typically thought of as separate, sequential phases in a halo's history; however, \textcolor{black}{it is possible for both types of star formation to take place simultaneously within the halo's merger tree. While we do not find any cases of Pop III and Pop II particles coexisting within a single halo, we do find cases of the two types of particles coexisting in different branches of the tree. The existence of such halos suggests that chemical enrichment history is complicated, as ejecta from the contemporaneous Pop III and Pop II particles mix into the final halo by proxy of halo merging}. The effective overlap between the star formation phases is demonstrated in Fig. \ref{fig:p3p2_overlap}, which shows the time difference between each Pop III supernova and the formation of the first Pop II star particle for each star forming halo in the simulation as identified in the final output. While most halos show little to no overlap between the phases, there are 7 halos in our sample with an overlap of over 100 Myr, with the longest overlap being nearly 300 Myr. The 4 most massive halos continue to form additional Pop II particles during this overlap phase. Figure \ref{fig:p3p2_overlap} also provides a measure of the maximum time scale for the Pop III phase of the halos in this simulation, which is about 300 Myr. This time scale is entirely dependent on a halo's merger history, and is thus subject to change if the simulation progressed further and more halos were allowed the time to merge.

\begin{figure*}[t]%
    \centering
        \includegraphics[width=0.94\textwidth]{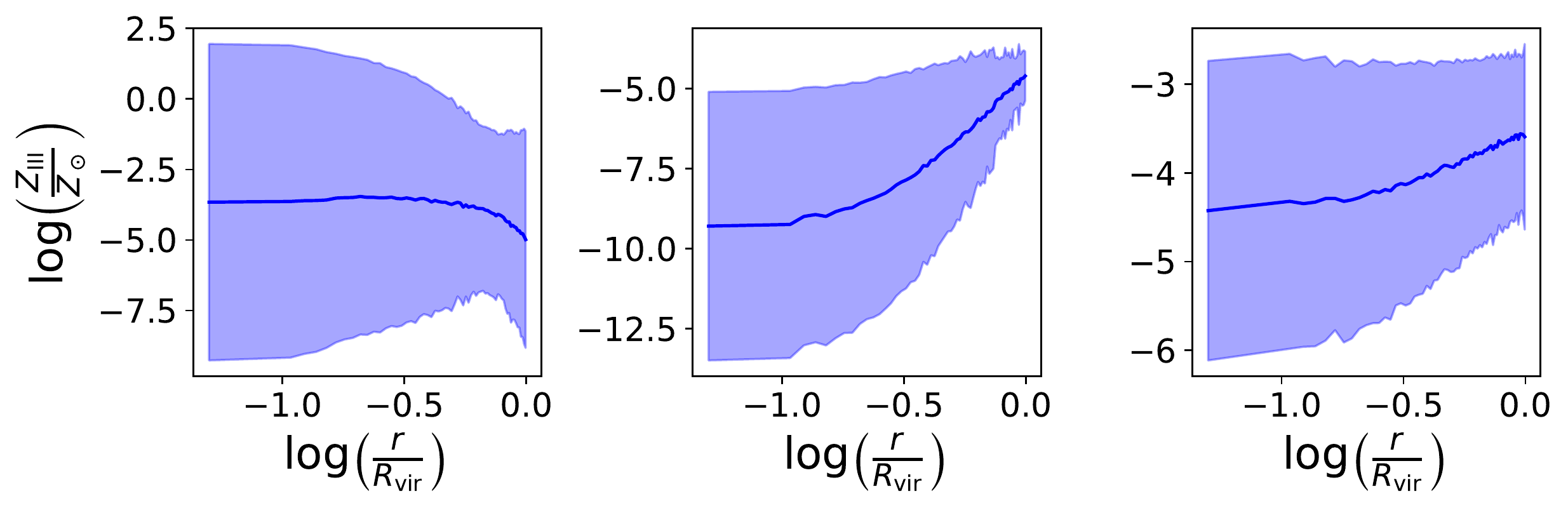}
    \caption{\textcolor{black}{Average Pop III metallicity profiles for internally enriched (\textbf{left}), externally enriched (\textbf{middle}), and pre-enriched (\textbf{right}) halos at the time they first exceed a mass-weighted average metallicity of $Z_\mathrm{crit}$. The halo counts are 136, 174, and 76, respectively. A profile for each halo of a given type is made with the same bins in $r/R_\mathrm{vir}$. The average and standard deviation in $\log{\left(Z_\mathrm{III}/Z_\odot\right)}$ are then taken over all the halos for each bin, resulting in the profile here. The blue line indicates the average, and the fill extends out to the standard deviation.}}
\label{fig:avg_profiles}
\end{figure*}


\textcolor{black}{An important result of Figures \ref{fig:p3p2_scatter} and \ref{fig:p3p2_overlap} is that externally enriched Pop II star formation is very rare in our simulation. We have already discussed that externally enriched halos primarily have low mass (i.e. Fig \ref{fig:f3_plot}), which means that star formation in these halos is unlikely. However, it is also possible that we lack sufficient resolution at the center of these halos to discern the complex metallicity distribution in potential star forming regions, which limits the performance of our criteria for selecting whether a newly-formed star particle will represent a Pop III star or a Pop II star cluster because only the properties of the densest cell are taken into account when distinguishing between the two types of particle (see Sec. \ref{sec:stars_and_feedback}). Figure \ref{fig:avg_profiles} shows the average Pop III metallicity profile at the time of enrichment for each of the three types of enriched halo discussed in this study: internally enriched, externally enriched, and pre-enriched. While the internally enriched and pre-enriched average profiles have metallicities above $Z_\mathrm{crit}$ at the center, the average externally enriched profile notably has metallicities $>4$ orders of magnitude below $Z_\mathrm{crit}$ at the center where stars would form. Similarly to Figure \ref{fig:exemplary_figure}, the externally enriched profile increases outwards from the center, and does not cross $Z_\mathrm{crit}$ until about $0.8\,R_\mathrm{vir}$. It should be emphasized that the time of enrichment does not necessarily align with the time of star formation in star forming externally enriched halos, so metal ejecta will likely have more time to mix inwards by the time star formation takes place. Whether this turbulent mixing is efficient enough to introduce variations in metallicity on scales smaller than the minimum cell size, 0.95 comoving pc (0.09 proper pc at the final output), to facilitate widespread Pop II star formation in externally enriched halos requires further study, but the large standard deviation in the externally enriched panel of Figure \ref{fig:avg_profiles} suggests that it is not unreasonable. The highly resolved externally enriched ``action halo" in \cite{doi:10.1093/mnras/stv1509} does achieve metallicities exceeding $Z_\mathrm{crit}$ (there, 10$^{-6}\,\,Z_\odot$) towards the center by the time gravitational collapse in the star forming core begins, but it is uncertain whether this is a common occurrence among externally enriched halos.}

\subsection{The Most Massive Halo}

\begin{figure}[t]
    \centering
    \includegraphics[width=0.47\textwidth]{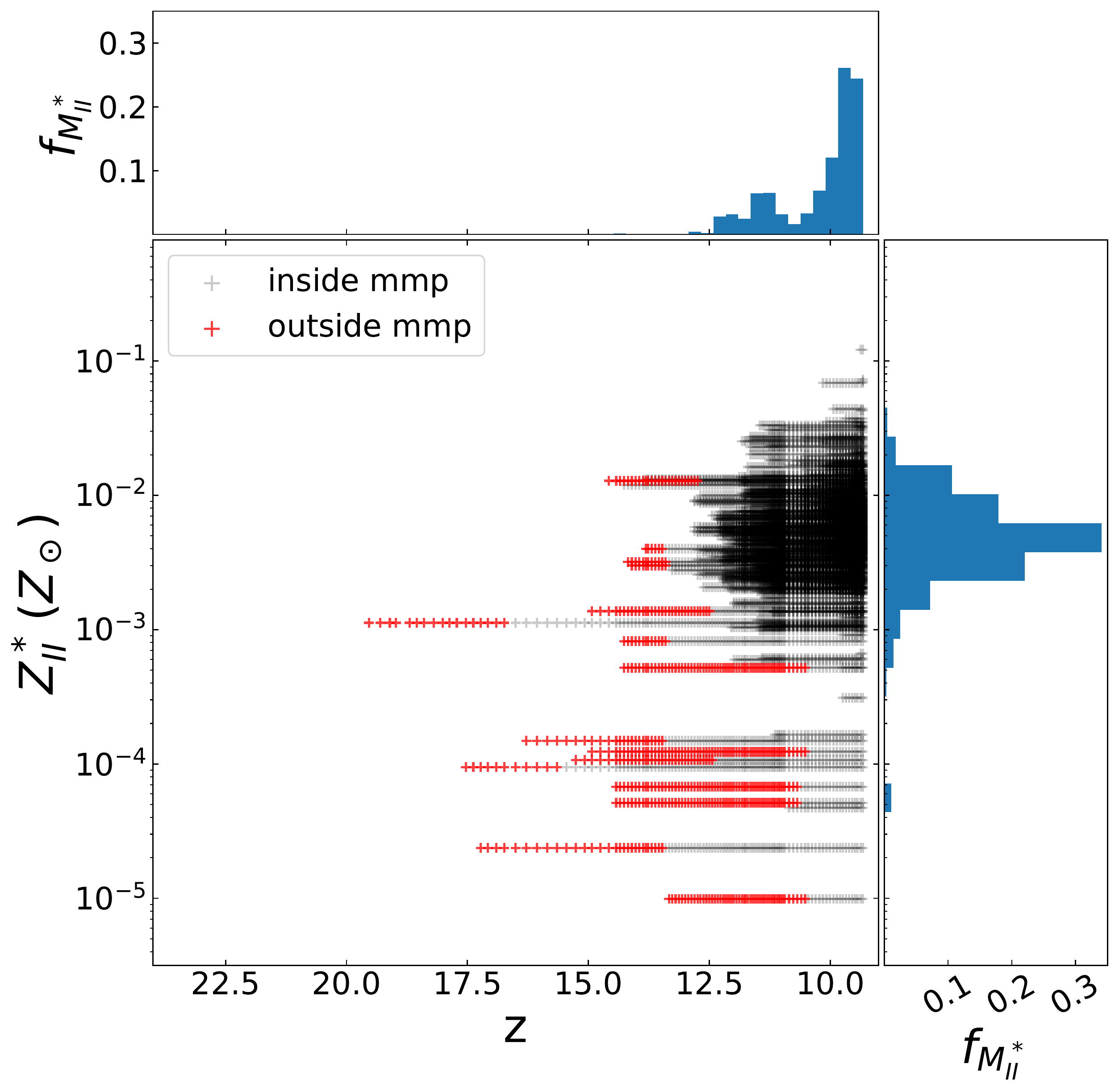}
    \caption{\textbf{Left:} Pop II formation history of the most massive halo in the simulation ($M_v = 3.71 \times 10^8 M_{\odot}$). Each data point represents a star particle in the merger tree. Red points correspond to star particles that are outside the most massive progenitor, while black points correspond to star particles that are inside. By the end of the simulation, the halo has a stellar mass of about $10^{6.3}\,M_\odot$. The discrete horizontal lines of points represent each individual star moving through time. \textbf{Right:} Stellar metallicity distribution of the halo at the final output. The variable, $f_{M_{II}^*}$, is the fraction of the halo's total stellar mass by the last output within each bin. The distribution peaks near $Z = 10^{-2.4}\,Z_\odot$. \textbf{Top:} Histogram of $f_{M_{II}^*}$ versus redshift of formation.}
    \label{fig:Zstar_most_massive}
\end{figure}

In order to assess the importance of external enrichment in the star formation history of the most massive halo in the simulation, a merger tree was created using Consistent Trees \citep{Behroozi2013}, and the most massive progenitor was tracked over time along with all of the stars that would eventually end up in the final halo. Figure \ref{fig:Zstar_most_massive} shows the Pop II star formation history and final stellar metallicity distribution function (MDF) for the most massive halo in the simulation ($M_v = 3.71 \times 10^8 M_{\odot}$), and distinguishes stars by whether or not they reside inside the most massive progenitor at a given redshift. Its first Pop II star particle forms around $z=20$ with a metallicity of about $10^{-3}\,Z_\odot$ outside of the most massive progenitor halo. Without double-counting due to stars forming within subhalos, the most massive halo has 5 separate Pop II star-forming progenitors. Stars that form through external enrichment are logged in the same way as was done for Figures \ref{fig:p3p2_scatter} and \ref{fig:p3p2_overlap} in the previous section; however, the most massive halo's history is devoid of externally enriched star formation. This is interesting because this halo contains 97\% of the simulation's Pop II stellar mass and is a major source of enrichment for nearby externally enriched halos.


\begin{figure}[t]
    \centering
    \includegraphics[width=0.47\textwidth]{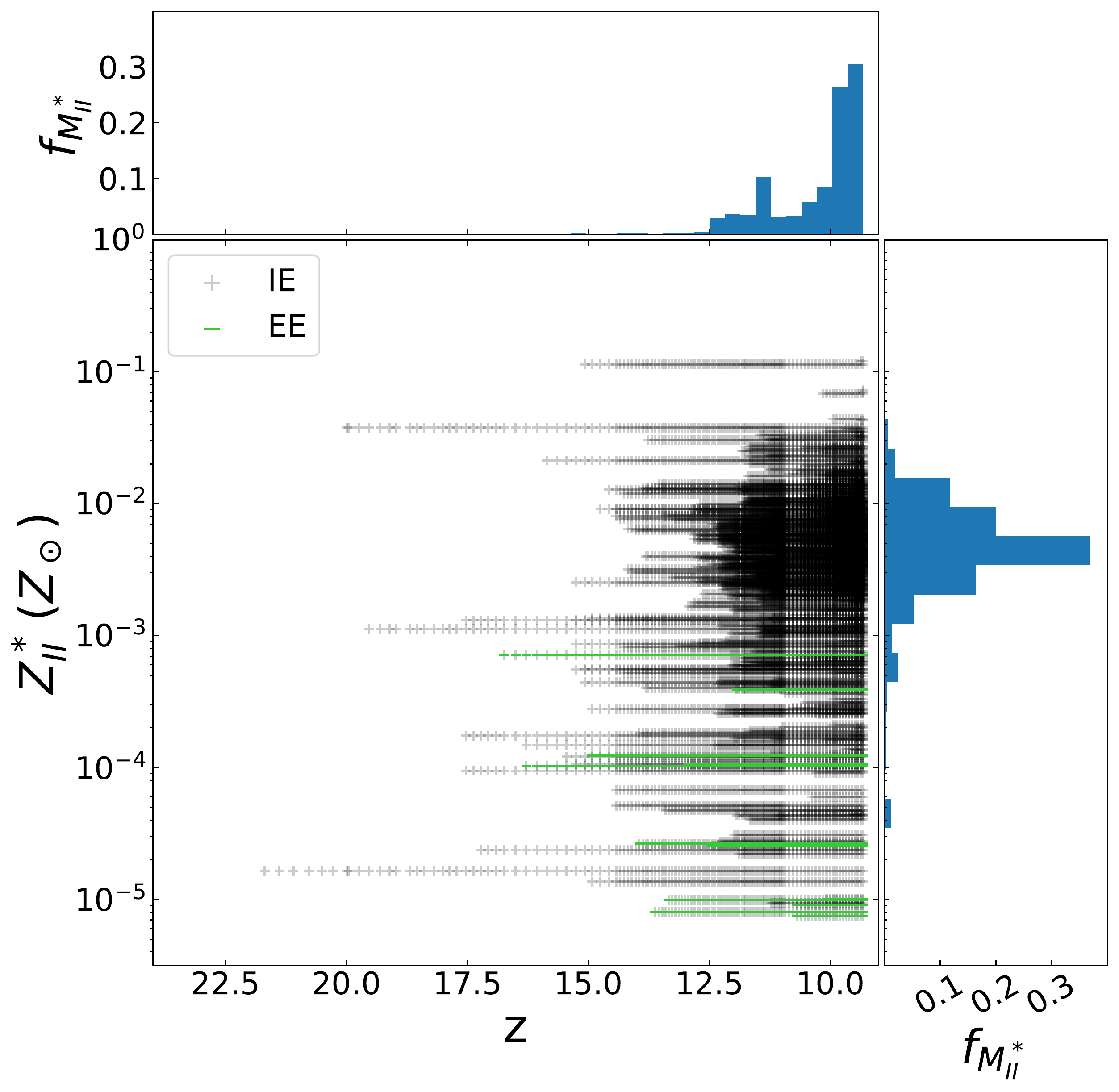}
    \caption{\textbf{Left:} Metallicity of all Pop II star particles in the simulation volume versus redshift. Each grey data point represents a star particle, with horizontally aligned points track the star over time. Stars that form in externally enriched halos are signified with green horizontal lines. \textbf{right:} Metallicity distribution function by mass fraction of Pop II star particles at $z=9.3$. Even though the critical metallicity for Pop II star formation is $Z_\mathrm{crit}=10^{-5.3}\,Z_\odot$, this distribution peaks near $Z=10^{-2.4}\,Z_\odot$. All of the particles that form through external enrichment have metallicity below $10^{-3}\,Z_\odot$. \textbf{Top:} Histogram of $f_{M_{II}^*}$ versus redshift of formation.}
    \label{fig:p2_mdf}
\end{figure}

The right panel of Fig. \ref{fig:Zstar_most_massive} shows the Pop II stellar MDF for the most massive halo. For comparison, Fig. \ref{fig:p2_mdf} shows the MDF for all Pop II particles in the simulation volume at the final output, and highlights those that form in externally enriched halos. The distribution peaks near $Z=10^{-2.4}\,Z_\odot$, and falls off dramatically below $Z=10^{-3}\,Z_\odot$ with a minimum value of $Z=10^{-5.1}\,Z_\odot$. There is no contradiction that the minimum metallicity is below $Z_\mathrm{crit}=10^{-5.3}\,Z_\odot$ for reasons discussed above.  All of the externally enriched star formation exists in the long tail of the distribution below $10^{-3}\,Z_\odot$. The MDF for the most massive halo looks very similar, as it contains most of the Pop II particles in the simulation, but it has fewer particles in the low-metallicity tail.

\section{Discussion}
\label{sec:discussion}

External enrichment phenomena are studied in \cite{jeon2017}, where external enrichment leads to extremely low-metallicity Pop II stars, and allows halos to form Pop II stars without ever hosting a Pop III star. This study corroborates that these halos can exist (i.e., Fig. \ref{fig:f3_plot}) in appreciable quantities, but most of the externally enriched halos identified in this simulation are not massive enough to support \textcolor{black}{star formation, and additionally do not have the metallicities towards the center exceeding $Z_\mathrm{crit}=10^{-5.3}$ required for Pop II star formation}. In agreement with \cite{doi:10.1093/mnras/stv1509}, the Pop II stars that do form through external enrichment form with very low absolute metallicity. However, we find such stars are absent from the merger tree of the most massive halo formed in this simulation. This finding could potentially be misleading, as this halo contains 97\% of the Pop II stellar mass summed over the entire simulation volume by the final output and is thus the only halo to achieve a stellar mass $\gtrsim10^5\,M_\odot$ in this sample. A much larger sample of star forming halos is needed to accurately assess the role of external enrichment in the formation of more massive galaxies. 

\cite{jeon2017} also comments on the transition time between Pop III and Pop II star formation, finding that the transition takes a few tens of Myrs. This is in agreement with our findings, for which we find an average transition time of 63 Myr. Additionally, we find that there is not always a hard line separating the two phases of star formation, as there can be a period lasting up to $\sim 100\,\mathrm{Myr}$, during which both phases can exist simultaneously within separate halos in the merger tree of the final halo.

The search for metal-poor stars has been the focus of many observational surveys. \cite{Abohalima_2018} compile an online database of stellar chemical abundance catalogs spanning searches from 1991 to 2016 that identified metal-poor stars located in the galactic halo, bulge, and dwarf galaxies in the Local Group. Currently, there are over 900 unique stars that have been identified with $[\mathrm{Fe/H}] \leq -2.5$. While stars with $[\mathrm{Fe/H}] < -4.0$ are exceptionally rare, surveys continue to identify stars with lower and lower chemical abundances. The current record-holder is the carbon-enhanced SMSS J0313-6708, which has $[\mathrm{Fe/H}] < -7.3$ (\cite{2014Natur.506..463K}). Despite its incredibly low iron abundance, the star's abundances of carbon and oxygen have upper bounds of $10^{-2.4}$ and $10^{-2.3}$ of solar, respectively. This places its total metallicity near the peak of Figure \ref{fig:p2_mdf}, which shows an MDF of all Pop II star particles in the simulation by the final output. We observe very broad ranges of metallicity for Pop II stars; the lower bound here is a combined artifact of 1) our choice of critical metallicity to form Pop II stars, and 2) the limited sample pool of Pop II stars in the simulation. 

\color{black}
Table 4 in \cite{Kirby2013} summarizes the MDFs for a collection of dwarf galaxies in the Local Group. The most massive halo in the simulation has a similar stellar mass and average Pop II stellar metallicity as the dwarf spheroidal (dSphr) Milky Way satellite, Ursa Minor, which has a stellar mass of $10^{5.73\pm0.20}\,M_\odot$ and an average iron abundance ratio, $\langle[Fe/H]\rangle$, of $-2.13\pm0.01$. The iron abundance ratio here is averaged over a sample of 190 stars that have been measured spectroscopically in the dwarf, and is thus a good indicator of the dwarf's total metallicity. Also similar is the dSphr, Sextans, with $M_\star=10^{5.84\pm0.20}\,M_\odot$ and $\langle[Fe/H]\rangle=-1.94\pm0.01$, averaged over 123 unique stars within. These two dwarf galaxies are characterized by their primarily old, metal-poor stellar populations, with the only significant bursts of star formation occuring early on in their lifetime (\citealt{Carrera2002, Bettinelli2018}).
\color{black}

In agreement with metal-poor DLA studies \citep{cooke2017, welsh2019}, the star-forming galaxies in our simulation have only had a few enriching events.  \cite{welsh2019} places an upper limit of $\lesssim 70$ enriching Pop III supernovae. We find our galaxies ($\sim 10^3-10^4\,M_\odot$) to have $\lesssim$ 20 enriching events, with the outlier most massive galaxy ($M_*\approx 10^{6.3}\,M_\odot$) displaying $>100$ events.  We also find agreement in that most of the enriching events were HNe or SNe in stellar mass range $10<M_*/M_\odot<40$, which is entirely due to the characteristic mass chosen for this simulation.  However, these DLA studies have not found evidence for highly energetic PISNe, which are included here.

\section{Conclusions}
\label{sec:conclusion}
\color{black}
We have analyzed the formation and chemical evolution history of a sample of halos with mass $10^{5.3} \leq M_\mathrm{vir}/M_{\odot} \leq 10^{8.6}$ derived from an Enzo AMR radiation hydrodynamic cosmology simulation which includes detailed models for Pop III and Pop II star formation and their chemical and radiative feedback within a 1 Mpc comoving box to a stopping redshift of z=9.3. The simulation is a re-run of \cite{wise2012b} with $1{,}000$ data outputs saved for subsequent analysis.  By the final redshift, 417 of the $1{,}864$ halos analyzed are chemically enriched to $\langle Z_\mathrm{total} \rangle > 10^{-5.3} Z_{\odot}$ where $\langle Z_\mathrm{total} \rangle$ is the halo's mass-averaged metallicity from both Pop III and II stellar enrichment. With our high time resolution we can distinguish between 3 enrichment pathways: (1) internally-enriched: Pop III stellar remnants within the halo's virial radius could have supplied the Pop III metals bound to it; (2) externally-enriched: Pop III stellar remnants within the halo's virial radius could not have supplied the Pop III metals bound to it, or contains no stellar remnants at all; (3) pre-enriched: a halo is born enriched with no Pop III stellar remnants. Based on our analysis, we can draw several conclusions:
\color{black}
\begin{enumerate}
\item Far from being the outlier, external enrichment is the dominant enrichment pathway that can provide enough metals to push the average metallicity of high redshift minihalos above the critical value required for Pop II star formation. Most halos that are enriched through this mechanism, however, are not massive enough to form stars. When external enrichment does trigger Pop II star formation, the resulting star particles have low metallicity.
\item Internal enrichment is the dominant pathway for halos forming Pop II stars in this simulation; \textcolor{black}{however, most Pop II star formation occurs in the most massive halo}.
\item Only a small percentage of enriched halos form pre-enriched, increasing from \textcolor{black}{1\% to 17.5\%} by the end of the simulation. 
\item The fraction of halos that are externally enriched increases over time, and the majority of these halos have virial mass below $10^{6}\,M_\odot$.  
\item \textcolor{black}{Pair-instability supernovae contribute the most to the enrichment of the IGM as a whole for our choice of primordial IMF, and are consequently the predominant supernova type contributing to the external enrichment of halos in spite of the fact that they only account for 8\% by number of the Pop III supernova events.}
\item \textcolor{black}{The average Pop III metallicity profile for externally enriched halos at the time of enrichment shows metallicities $>4$ orders of magnitude below the critical metallicity for Pop II star formation, $Z_\mathrm{crit}$, at the center, with values increasing outwards until they reach $Z_\mathrm{crit}$ at about $0.8\,R_\mathrm{vir}$. In contrast, the average profile for internally enriched halos has metallicities above $Z_\mathrm{crit}$ throughout, with values decreasing outwards. The profile for pre-enriched halos also has metallicities above $Z_\mathrm{crit}$ throughout, but it is flatter than the other two cases, increasing outwards by less than an order of magnitude.}
\item The line between the Pop III and Pop II star formation phases in a merger tree is blurry, as a halo that is currently in the Pop II phase can merge with a halo that is still in the Pop III phase. The period of overlap, where both Pop III and Pop II star formation takes place within a merger tree, typically lasts around 100 Myr for the halos in this simulation.
\end{enumerate}

We have found that during early star formation, the metal enrichment process is not exclusively local to the host halo, and must include the surrounding environment. The region that must be included will likely depend on many factors, e.g., halo mass, halo number density, baryon density between halos and enriching events, temperature, and the particular type of supernova. A future study may be able to study these variables more precisely to determine their effect on the enrichment process.  

\bigskip
\noindent {\em Acknowledgements:} This research was supported by National Science Foundation CDS\&E grant AST-1615848 to M.L.N. and B. S. ~~A.W. gratefully acknowledges support from a MPS Graduate Research Supplement for Veterans (MPS-GRSV) fellowship to the above award. JHW is supported by NSF grants AST-1614333 and OAC-1835213 and NASA
grants NNX17AG23G and 80NSSC20K0520. BWO acknowledges support from NSF  grants  PHY-1430152,  AST-1514700, AST-1517908, and  OAC-1835213,  by  NASA grants NNX12AC98G and NNX15AP39G, and by HST-AR-13261 and HST-AR-14315. The simulations
were performed using ENZO on the Blue Waters supercomputer operated
by the National Center for Supercomputing Applications
(NCSA) with PRAC allocation support by the NSF (award
number ACI-0832662). Data analysis was performed on the
Comet supercomputer operated for XSEDE by the San Diego
Supercomputer Center. Computations and analysis described in this work were performed using the publicly-available Enzo \citep{Enzo2014, Enzo2019}, yt \citep{YT}, and ytree \citep{ytree} codes, which is the product of a collaborative effort of many independent scientists from numerous institutions around the world.

\bibliographystyle{aasjournal}
\bibliography{bib}
\end{document}